\documentclass[journal]{IEEEtran}

\usepackage{subfigure}
\usepackage{graphicx}
\usepackage{algorithm,algorithmic,epsfig}
\usepackage{amsmath}
\usepackage{amsfonts}
\usepackage{verbatim}
\usepackage{array}
\usepackage{soul}
\usepackage{color}
\usepackage{authblk}
\usepackage{framed}
\usepackage{balance}
\usepackage{picins}
\usepackage{filecontents}
\usepackage{afterpage}
\usepackage{fancyhdr}

\makeatother

\fancypagestyle{IEEE_Green_open_access_footer}{
  \fancyhf{}
  \fancyhead[C]{\footnotesize \copyright2019 IEEE. Personal use of this material is permitted. Permission from IEEE must be obtained for all other uses, in any current or future media, including reprinting/republishing this material for advertising or promotional purposes, creating new collective works, for resale or redistribution to servers or lists, or reuse of any copyrighted component of this work in other works. DOI: 10.1109/MCOM.2019.1800611}     

}

\begin{document}

\title{Toward Understanding Crowd Mobility \\in Smart Cities through the Internet of Things}

\author[1]{G{\"u}rkan~Solmaz}
\author[2]{Fang-Jing~Wu}
\author[1,3]{Flavio~Cirillo}
\author[1]{Ern\"o Kovacs}
\author[4]{\\Juan~Ram{\'o}n~Santana}
\author[4]{Luis~S{\'a}nchez}
\author[4]{Pablo~Sotres}
\author[4]{Luis~Mu{\~n}oz}
\affil[1]{NEC Laboratories Europe, Heidelberg, Germany}
\affil[2]{TU Dortmund University, Dortmund, Germany}
\affil[3]{University of Napoli Federico II, Napoli, Italy}
\affil[4]{University of Cantabria, Santander, Spain}
\maketitle
\thispagestyle{IEEE_Green_open_access_footer}

\begin{abstract}
Understanding crowd mobility behaviors would be a key enabler for crowd management in smart cities, benefiting various sectors such as public safety, tourism and transportation. This article discusses the existing challenges and the recent advances to overcome them and allow sharing information across stakeholders of crowd management through Internet of Things (IoT) technologies. The article proposes the usage of the new {\em federated interoperable semantic IoT platform} (FIESTA-IoT), which is considered as ``a system of systems''. The platform can support various IoT applications for crowd management in smart cities. In particular, the article discusses two integrated IoT systems for crowd mobility: 1) {\em Crowd Mobility Analytics System}, 2) {\em Crowd  Counting  and  Location  System}  (from the SmartSantander testbed). Pilot studies are conducted in Gold Coast, Australia and Santander, Spain to fulfill various requirements such as providing online and offline crowd mobility analyses with various sensors in different regions. The analyses provided by these systems are shared across applications in order to provide insights and support crowd management in smart city environments.

\end{abstract}

\IEEEpeerreviewmaketitle

\section{Introduction}
\label{Introduction}
Sustainable development of cities is a major global challenge as more than half of the world population is living in urban areas. The smart city concept allows optimizing services for urban areas because or as a result of the advancement of the new technologies ranging from very small devices to big data centers. These technologies can be considered in the context of IoT, where many objects, devices, machines, and data centers are connected. The usage of IoT technologies for {\em crowd management} in urban environments is promising for the future of smart cities.

IoT technologies can enable many improvements for crowd management, which spans sectors such as transportation services (e.g., operating public transport or directing pedestrian traffic), public safety (e.g., detection of fighting incidents), and tourism (e.g., event management for enhanced visitor experience). For instance, movement behaviors of crowds may indicate situations such as traffic congestion, emergency incidents, and panic situations during certain events such as large gatherings in city squares.

\begin{figure}
\centering
\includegraphics[clip, trim={3.1cm 1cm 4.5cm 1cm}, width=0.9\columnwidth]{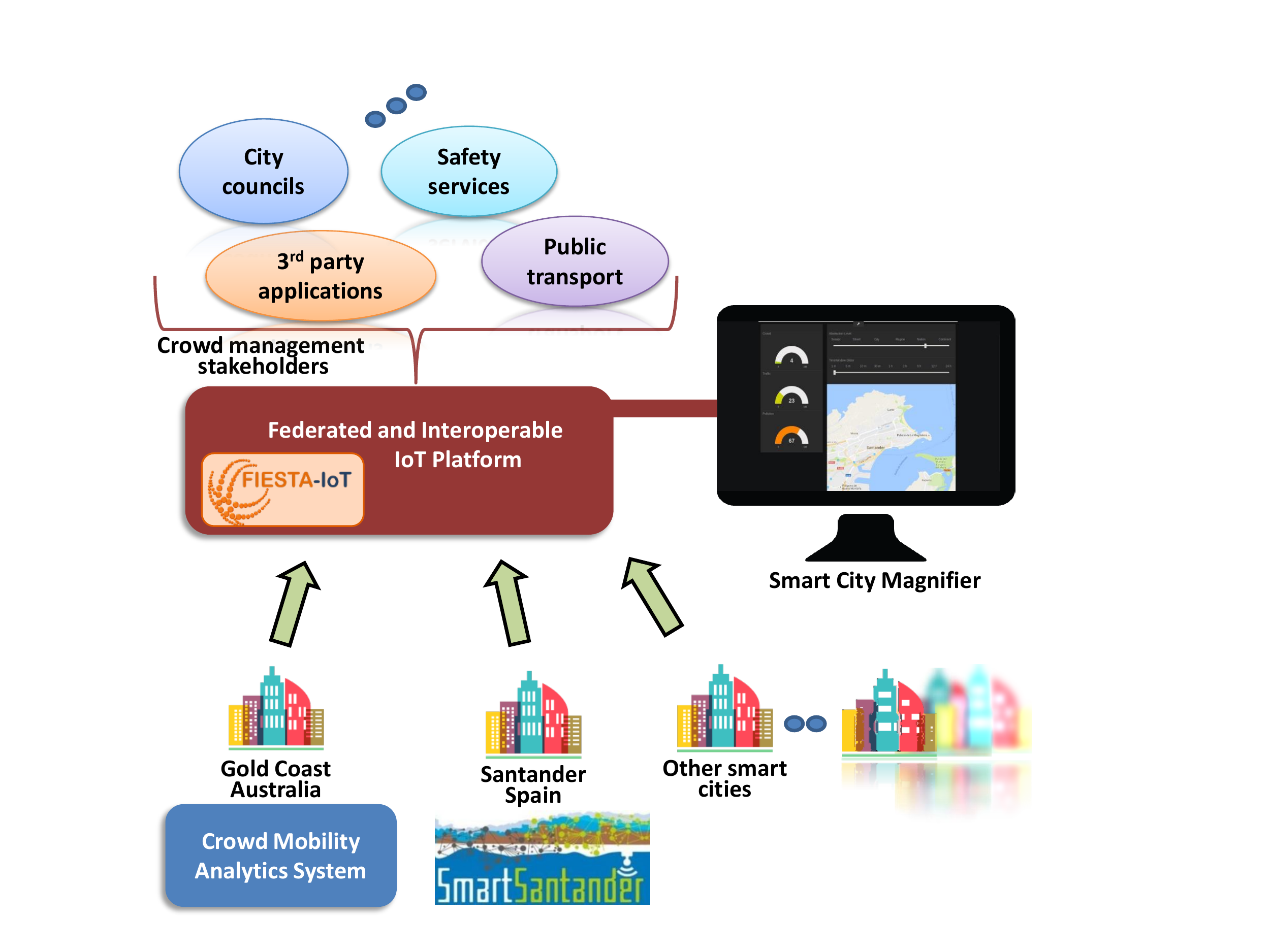}
\caption{Federated and interoperable IoT platform supporting crowd management stakeholders of smart cities.} \label{Fig:IoTPF}
\end{figure}

While cities aim to achieve smart urban services, new challenges arise due to the limitations and deficiencies of the current systems and technologies in terms of \emph{scalability} of the connected systems, \emph{information transparency} between different systems (i.e., semantic interoperability) or stakeholders, \emph{data federation}, and \emph{information privacy}. When mobility information must be shared across multiple stakeholders, a proprietary infrastructure cannot fulfill all the different requirements that they impose. For example, some of the stakeholders expect real-time mobility monitoring service for event detection while others require historical mobility data analytics to analyze efficiency of services in different urban environments (e.g., train station, stadium, city square). It is very difficult to re-design a one-size-fit-all IoT system when new requirements arise for various environments or different time periods. A better solution is to provide ``a system of systems'' in which a new service can be easily developed or setup to handle any new requirement by leveraging existing technologies and infrastructure. To make such a system of systems useful, semantic models based on an appropriate ontology are needed for transparently exchanging data, analytics results, and allowing to share new insights from different crowd management applications. The federation of data, results, and learned insights is the key technical enabler to understand the crowd mobility behaviors in a smart city. Finally, privacy preservation is a problem of utmost importance for smart cities. While various data from vast deployment of sensors travel through the IoT systems, preserving privacy at a level closer to the data contributors (providers) is an important challenge.

This article describes the recent advances in IoT for understanding crowd mobility in smart cities. The {\em federated and interoperable semantic IoT} (FIESTA-IoT) platform for smart cities is introduced for the specific perspective of crowd management applications. Fig.~\ref{Fig:IoTPF} illustrates the outlook of the smart city applications leveraging the smart city platform for sharing information across various stakeholders. While the platform is currently in use for several smart city testbeds, the article focuses on two IoT systems for crowd mobility, namely {\em Crowd Mobility Analytics System} (CMAS) and {\em Crowd Counting and Location System} (CCLS) and discusses the aspects related to the aforementioned limitations.

Two pilot studies are conducted in Gold Coast, Australia and Santander, Spain, where various sensors are deployed in urban areas. The first pilot study uses CMAS in Gold Coast for a medium-scale smart city deployment. The requirements of the pilot include analyzing heavy or light pedestrian traffic at streets with or without vehicles. The second pilot study uses CCLS in an indoor market in Santander. The requirements include detecting people (crowd size) and locating their positions at public buildings of a city and other critical infrastructures. In both pilots, data anonymisation limits tracking devices for long time periods. On the other hand, online and offline analytics information needs to be shared across various stakeholders such as city councils and visualized in several interfaces using IoT technologies and infrastructure to provide insights for crowd management in smart cities.

\section{Crowd Mobility Analytics using the Smart City Platform}
\label{Systems}

\subsection{Federated and Interoperable IoT Platform}

Smart city data is often gathered by solutions where dedicated networks of sensors or data sources produce observations to be consumed by specific applications. The systems usually differ from each other, serving for distinct purposes, and they are mostly not interoperable~\cite{zanella2014smartcities,yannuzzi2017fog}. In this regard, creating crowd management services that harness the abundant data from a smart city (e.g., environmental data, road traffic information) would require either ad-hoc integration or creation of new systems. This situation raises a new requirement of an integrated ``system of systems'' or ``container of systems''.

\begin{figure}
\centering
\includegraphics[clip, trim={2.8cm 0cm 4cm 0cm},width=0.96\columnwidth]{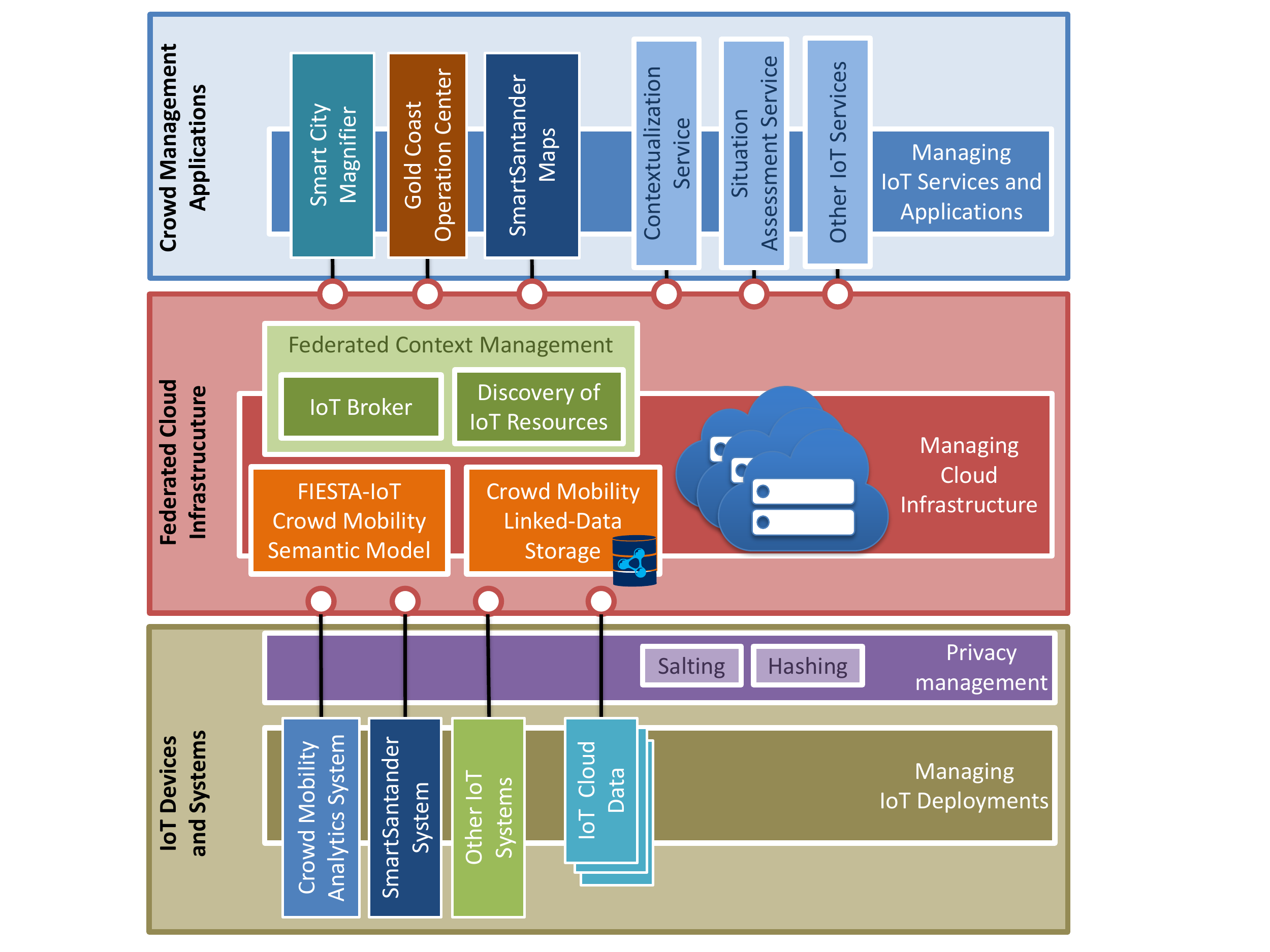}
\caption{Crowd mobility-based instantiation of the federated and interoperable IoT platform.}
\label{Fig:Architecture}
\end{figure}

To overcome this challenge, we propose a crowd mobility-based instantiation of the FIESTA-IoT platform~\cite{Lanza2018} and provide semantic interoperability from IoT deployments to the services (shown in Figure \ref{Fig:Architecture}). The heterogeneous IoT deployments on the \emph{IoT Devices and Systems} (bottom layer) are integrated to the Cloud and data is anonymised with salting and hashing. In this layer, in addition to the two crowd mobility systems (CMAS and CCLS), IoT Cloud Data from external platforms can be connected. Furthermore, there exist other IoT FIESTA-IoT systems that can be leveraged. Currently more than 5000 sensors (from 11 integrated testbeds~\cite{Sanchez2018}) report environmental data (e.g., temperature, humidity, illuminance, noise level), road traffic information (e.g., vehicle speed, traffic intensity), car and bike parking spots, estimated arrival times of buses, and smart building information (e.g., human occupancy, power consumption). 

At the \emph{Federated Cloud Infrastructure} (middle layer), the data from the bottom layer is modelled using the FIESTA-IoT Semantic Model and stored in the Linked-Data Storage. In particular, the semantic model for crowd mobility data is described in Section \ref{ontology}. The data in the Cloud infrastructure is accessible through the Federated Context Management which exposes NGSI and SPARQL interfaces. Our open source IoT Broker (Aeron Broker) component provides scalable federation for the context management, whereas IoT Discovery (NEConfMan) enables easy registration and discovery of resources with features such as geo-discovery. 

The crowd management-related IoT data is harnessed by {\em Crowd Management Applications} (top layer) which contain IoT services provided by the platform and crowd mobility applications. These services enhance the crowd mobility data through reasoning by aggregating the semantic data and assessing the situations related to physical objects (i.e., Contextualization Service) at different levels of abstraction such as buildings level or street level. Assessment of the situations can be performed through; a) pre-defined thresholds, b) anomaly detection, c) time-series analysis, d) artificial intelligence. The obtained situations are displayed on the dashboard in Figure \ref{Fig:IoTPF}, named {\em Smart City Magnifier}, which reports alerts regarding traffic status, crowd flows, critical events (e.g., fire bursting), and so on. Moreover, crowd mobility applications such as {\em Gold Coast Operation Center} and {\em SmartSantander Maps} receive the results (generated by CMAS, CCLS or other IoT services) from the Cloud and provide visualizations.

\subsection{Crowd Mobility Semantic Model}
\label{ontology}

\begin{figure*}[!t]
\centering
\includegraphics[clip, trim={0cm 1cm 0cm 0cm}, width=2\columnwidth]{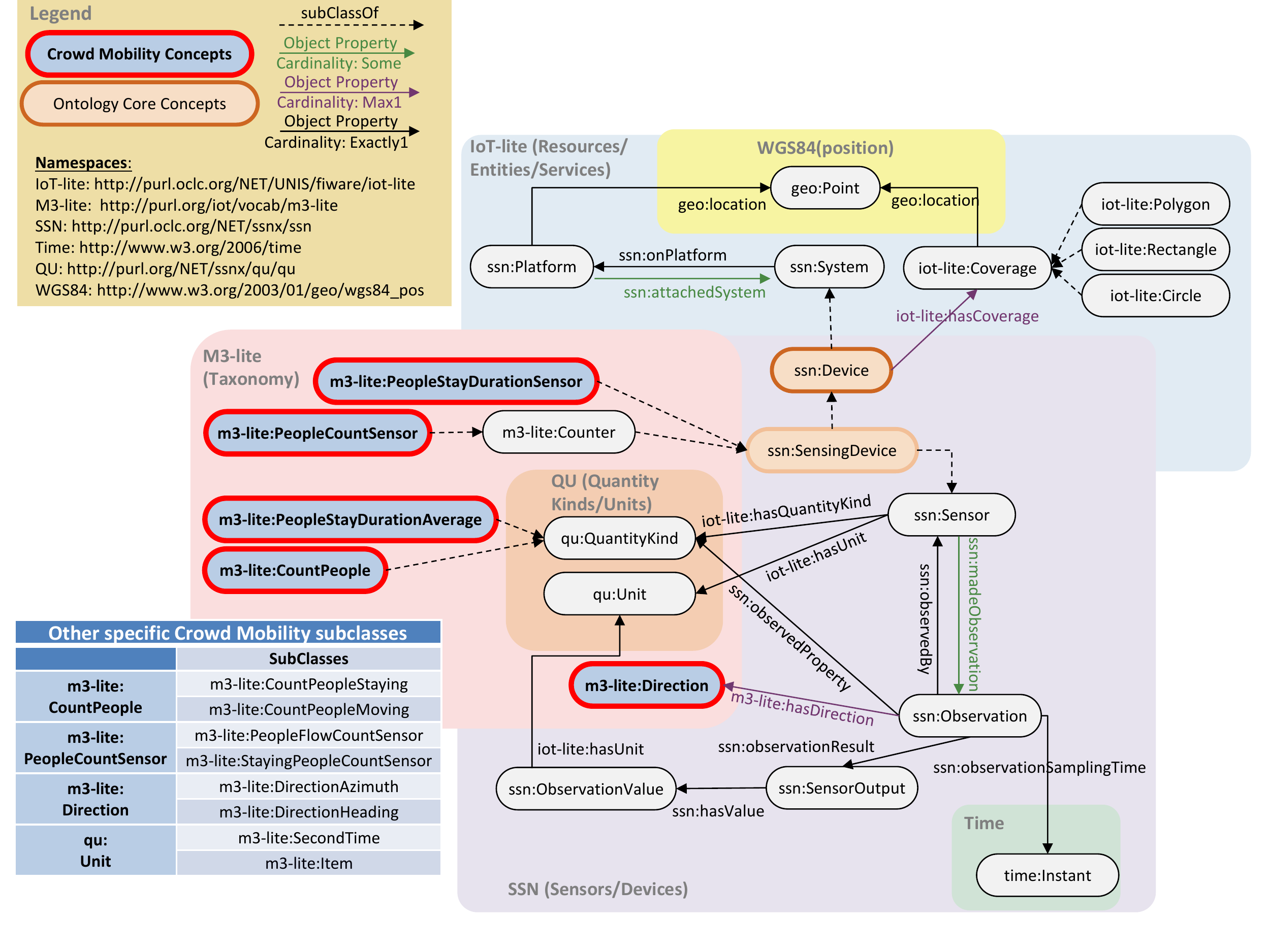}
\caption{Modeling crowd mobility information based on FIESTA-IoT ontology.} \label{Fig:Ontology}
\end{figure*}

In order to provide seamless interoperability and information transparency from IoT systems to the crowd management applications, the crowd mobility outcomes are semantically annotated following the FIESTA-IoT ontology~\cite{agarwal2016fiestaontology} as shown in Fig.~\ref{Fig:Ontology} (with a stress on the specific taxonomy of M3-lite 
for crowd mobility).

Rich and complex knowledge is represented with an ontology as things are connected to each other through relationships. Things are not identified as individuals, but as classes of individuals. Moreover, a class might have sub-classes. For example, peopleCounterX is an instance of \textit{PeopleCountSensor} class which is a subclass of \textit{Counter} (see Fig.~\ref{Fig:Ontology}). The classes can be defined and described in taxonomies and an ontology may use classes from different ontologies or taxonomies. Relationships between classes are known as properties and it is possible to define properties' cardinality. Each class and property used in an ontology is uniquely identified by a namespace prefix and the class or property specific name. For example, \textit{m3-lite:PeopleCountSensor} is a class defined in the M3-lite ontology. For the sake of readability, in this paragraph we are omitting the namespace prefix while they are shown with prefix in Fig.~\ref{Fig:Ontology}.

The core concept is the \emph{SensingDevice}, representing a sensor that produces \emph{Observation}, which is a measurement (or computation) of a phenomenon related to an object happened at a specific \emph{Instant}. For example, a crowd mobility detector can be seen as a \emph{Device} composed of multiple \emph{SensingDevices}. In this sense, such a detector can have one \emph{PeopleFlowCountSensor} and one \emph{StayingPeopleCountSensor}, which are subclasses of \emph{PeopleCountSensor}. The \emph{Observation}(s) is expressed with a \emph{QuantityKind} having a \emph{Unit}. Following our example, the \emph{QuantityKind} associated to the data generated by the \emph{PeopleFlowCountSensor} is \emph{CountPeopleMoving} (subclass of \emph{QuantityKind}) with \emph{Item} as its \emph{Unit} and with the \emph{Direction} property expressed either in geodetic \emph{DirectionAzimuth} or as a generic \emph{DirectionHeading}. The directions start from the \emph{Point} that is the \emph{location} of the physical \emph{Platform}. \emph{Platform} is meant as the supporting dock to which the \emph{Device} is attached. The \emph{StayingPeopleCountSensor} generates \emph{CountPeopleStaying} values expressed in \emph{Item}. The system also consists of \emph{PeopleStayDurationSensor} that generates \emph{PeopleStayDurationAverage} values measured in \emph{SecondTime}. Each \emph{SensingDevice} might have a \emph{Coverage}, specified either as \emph{Polygon/Rectangle/Circle} or as a simple \emph{Point}. This indicates the geographic extent of the \emph{Observation}.

\subsection{Integrated IoT Systems}
\label{CrowdMobilitySystems}
\subsubsection{Crowd Mobility Analytics System}
\label{CEMASystem}
The CMAS (extended from our system in~\cite{Wu-2018-MOBISYS}) is integrated with the platform via semantic annotation of the outcome. The developed system consists of Wi-Fi sniffers, stereoscopic cameras, IoT gateways, and data analytics modules. The Wi-Fi sniffers are capable of capturing wireless probes broadcasted by mobile devices. Based on the captured Wi-Fi probes, the system can count the mobile devices in these sensing areas. The cameras are co-located with specific Wi-Fi sniffers deployed at the dedicated {\em calibration choke points}. A built-in people counting software runs in the cameras. Both Wi-Fi device detection and people counting results are reported to to the Cloud, where data analytics modules reside, through the IoT gateways. Three analytics modules are developed: {\em crowd estimation, people flows, and stay duration}. The {\em crowd estimation} module outputs number of people by correlating the stereoscopic camera counts and the number of Wi-Fi enabled devices at the calibration points. Based on the correlation between the two data modalities, the calibration of the data analytical results are applied in other sensing areas without cameras. The module monitoring {\em people flows} infers crowd movement in these areas. Finally, the {\em stay duration} module estimates the waiting times and the number of waiting people. All analytics results are exported to the Federated Cloud Infrastructure so the crowd analytics results are discoverable and available for applications in the smart city platform.

\subsubsection{Crowd Counting and Location System}
\label{SMSSystem}
Different from CMAS, CCLS aims at analysing crowd behaviour in public buildings of a city, as well as critical infrastructures. The system relies in the analyses of IEEE802.11 frames to discover devices in the surroundings of the deployment, normally within the monitored areas. Similar to CMAS, the deployed nodes capture ``Probe Request'' frames sent by smartphones, which include a Wi-Fi interface in ``active search'' mode, incorporated in most of them. However, CCLS does not only aim at detecting people, but also aims at locating them. For this, the system stores the RSSI and sequence number from the captured frames. It is possible to locate people by processing this information using RSSI-based algorithms. All the post-processing is performed in an edge server, where all the measurements are sent after the corresponding anonymisation techniques are applied.
Once the anonymised raw measurements are analyzed and the counting and location analytics applied over them (i.e., the estimated crowd size and positions are obtained), these observations are semantically annotated and pushed to the Federated Cloud Infrastructure. For the semantic modelling, each crowd estimator is modelled as an \emph{PeopleCountSensor}, with a specific \emph{Coverage} (representing the area to which the estimations apply), that generates \emph{CountPeople} observations expressed in \emph{Item}.
\subsection{Privacy Considerations}
One of the essential requirements is dealing with tracked devices' privacy. Nowadays, privacy is one of the major public concerns. In this sense, data protection laws have to be observed when handling data that could be personal. Quite restrictive rules apply in most countries of the world, being the countries from the European Union (EU) some of the most restrictive ones. These rules are recently updated through the General Data Protection Regulation (GDPR)~\cite{regulation2016regulation} enforcement.

The Wi-Fi sensors in CMAS and CCLS deal with MAC addresses, which are considered personal data under the new EU regulation. As it is stated in the GDPR~\cite{regulation2016regulation}, ``The principles of data protection should apply to any information concerning an identified or identifiable natural person''. Therefore, Wi-Fi-based tracking services in public or private spaces can be performed only if the service obtains the user's opted-in permission, or data is anonymised in such manner that the user is no longer identifiable, as mentioned in the 26\textsuperscript{th} article from the aforementioned regulation.
The Article 29 Working Party, recently replaced by the European Data Protection Board (EDPB), is in charge of analysing the compliance of the privacy rules. In a document released to analyze the ePrivacy regulation compliance with GDPR~\cite{WP247}, the Data Protection Working Party states that Wi-Fi tracking can only be performed either if there is consent or the personal data is anonymised. Within the same document, four conditions are mentioned for the latter case to be compliant with the GDPR:
\begin{itemize}
    \item The purpose of the data collection from terminal equipment is restricted to mere statistical counting.
    \item The tracking is limited in time and space to the extent strictly necessary for this purpose.
    \item The data will be deleted or anonymised immediately afterwards.
    \item There exist effective opt-out possibilities.
\end{itemize}

Considering that user's permission request is impossible to obtain in normal conditions within the subject of the experimentation, the only option is to anonymise data regarding to MAC addresses. Thus, experimentation security measures must be undertaken to address both, data integrity and anonymisation. Therefore, any type of experimentation or service provision must take into account this concern, which is usually underestimated by system developers.   

CCLS in Santander is based on the Spanish Personal Data Protection Laws and the Spanish Law Protection Office recommendations for data anonymisation~\cite{agencia}. 
The recommendation consists on the use of a cryptographic hash function with randomly generated hash keys. More precisely, the HMAC protocol, which provides such mechanisms, is recommended. In the SmartSantander deployment, we implement the HMAC algorithm along with the SHA256 hashing function, with a 12-bytes randomly generated key. Finally, in order to ensure a non-reversible process, this implementation also comprises a procedure to destroy and renew the key during specific session periods. For CMAS in Gold Coast, the hashed and salted Wi-Fi probe data is sent to the Cloud. The stereoscopic cameras do not record video or perform face detection. The cameras simply count the passage of people through predefined lines at the choke points. The outputs of the camera are people count-in and -out values. The main drawback of this procedure is the limitation of tracking devices throughout long periods (as in~\cite{de2013unique}) or longer travels within the city, but it is the price that must be paid to meet the privacy requirements.

\section{Pilot Studies in Australia and Spain}
\label{PilotDescription}

\subsection{Pilot Deployment in Gold Coast}
\label{CEMASystemAdvanced}

\subsubsection{Pilot Setup}

The deployments in Gold Coast include 17 Wi-Fi sensors and 2 stereoscopic cameras. The Wi-Fi sensors are custom-built devices for outdoor deployments. Two cameras are used at the calibration choke points, where there is a camera and a Wi-Fi sensor deployed together. The cameras are the Hella Advanced People Sensor APS-90E 
deployed at a height about 3.6 meters. Each camera is configured to capture the entire choke point for accurate counting. 

The deployments target two regions. These sensors deployed in these areas are considered as {\em Cluster 1} for (expected) heavy pedestrian traffic and {\em Cluster 2} for light traffic places. Each cluster has a stereoscopic camera for the calibration. The collected data is sent to the Cloud where two virtual machines are created for the clusters. Clustering the areas allows applying CMAS to city-scale by sharing the raw data load.

\subsubsection{Pilot Operation}
The pilot study activities started in September 2017 and CMAS has been in use starting from November 2017. Various types of pilot tests are conducted on the field during the operation of the pilot. Manual counting is performed using video footages taken from different deployment areas. In comparison to manual counting, the cameras provide an accuracy between 88\% and 98\%, which mainly depends on the weather and lightning conditions. Furthermore, field tests for heavy and light traffic areas resulted in 93\% and 89\% crowd size accuracy compared to manual counting. The results obtained from outside the choke points give further confidence to treat stereoscopic camera results as {\em near ground truth} as proposed in~\cite{Wu-2018-MOBISYS}.

Gold Coast pilot successfully tests the crowd mobility analytics services by leveraging federation of clusters and interoperability using the semantic model to share the results with stakeholders. This shows that similar systems can be developed and leveraged by future crowd management applications using the smart city platform.

\begin{figure}
\centering
\includegraphics[width=0.9\columnwidth]{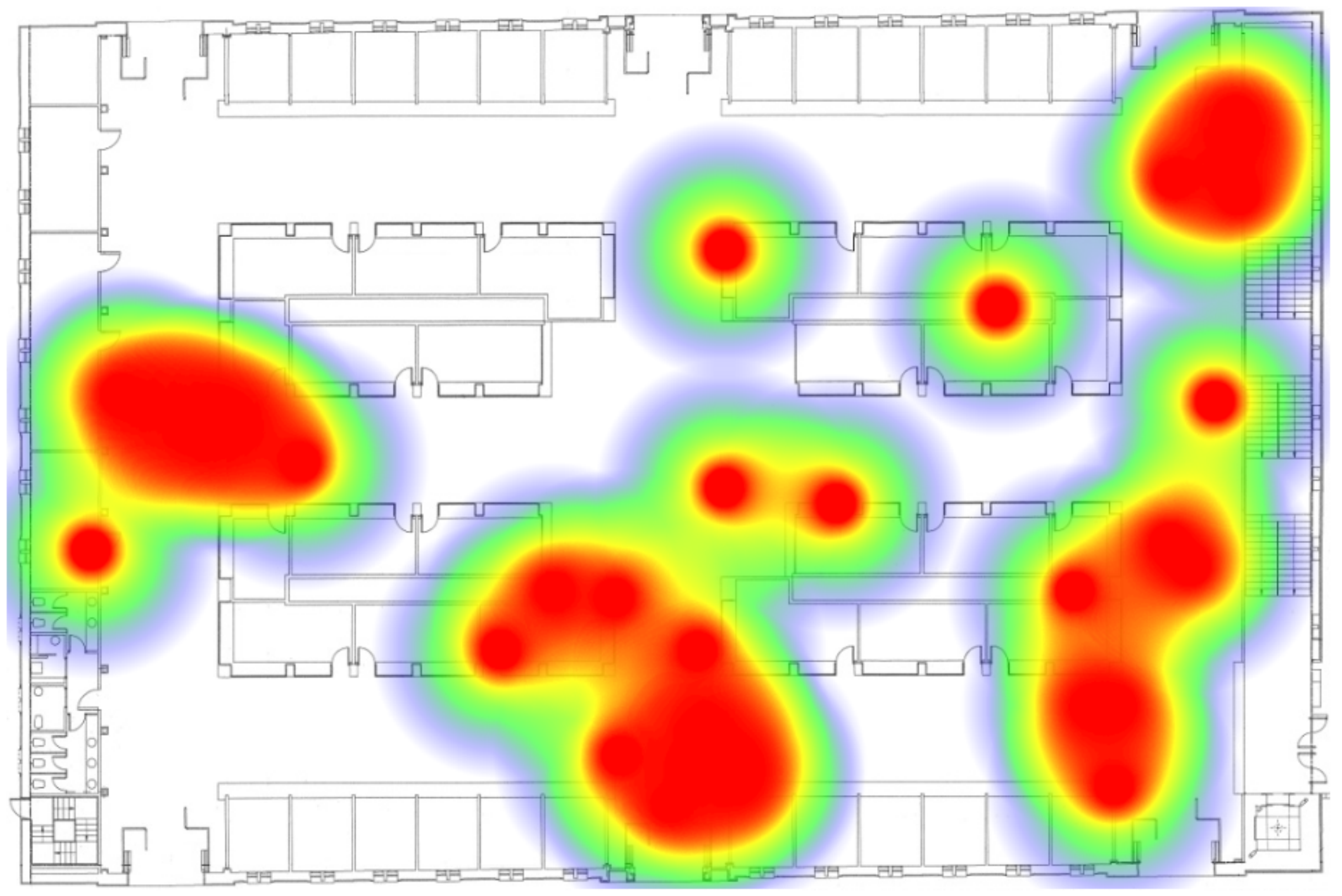}
\caption{Heatmap from Mercado del Este in Santander.} \label{Fig:MercadoDelEsteHeatmap}
\end{figure}

\begin{figure*}
\begin{framed}
\centering

\begin{tabular}{cc}
\includegraphics[width=0.44\linewidth]{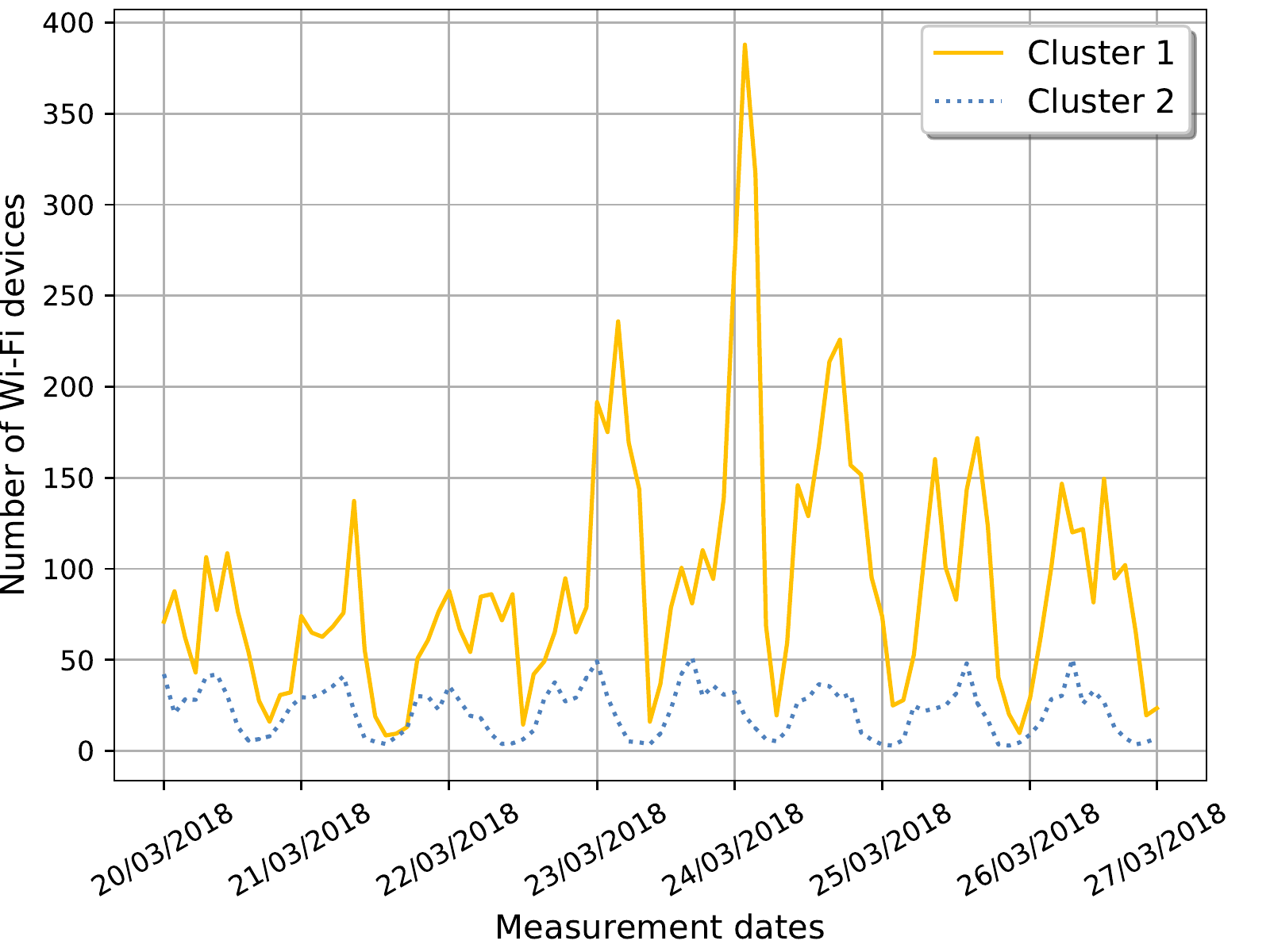} &
\includegraphics[width=0.44\linewidth]{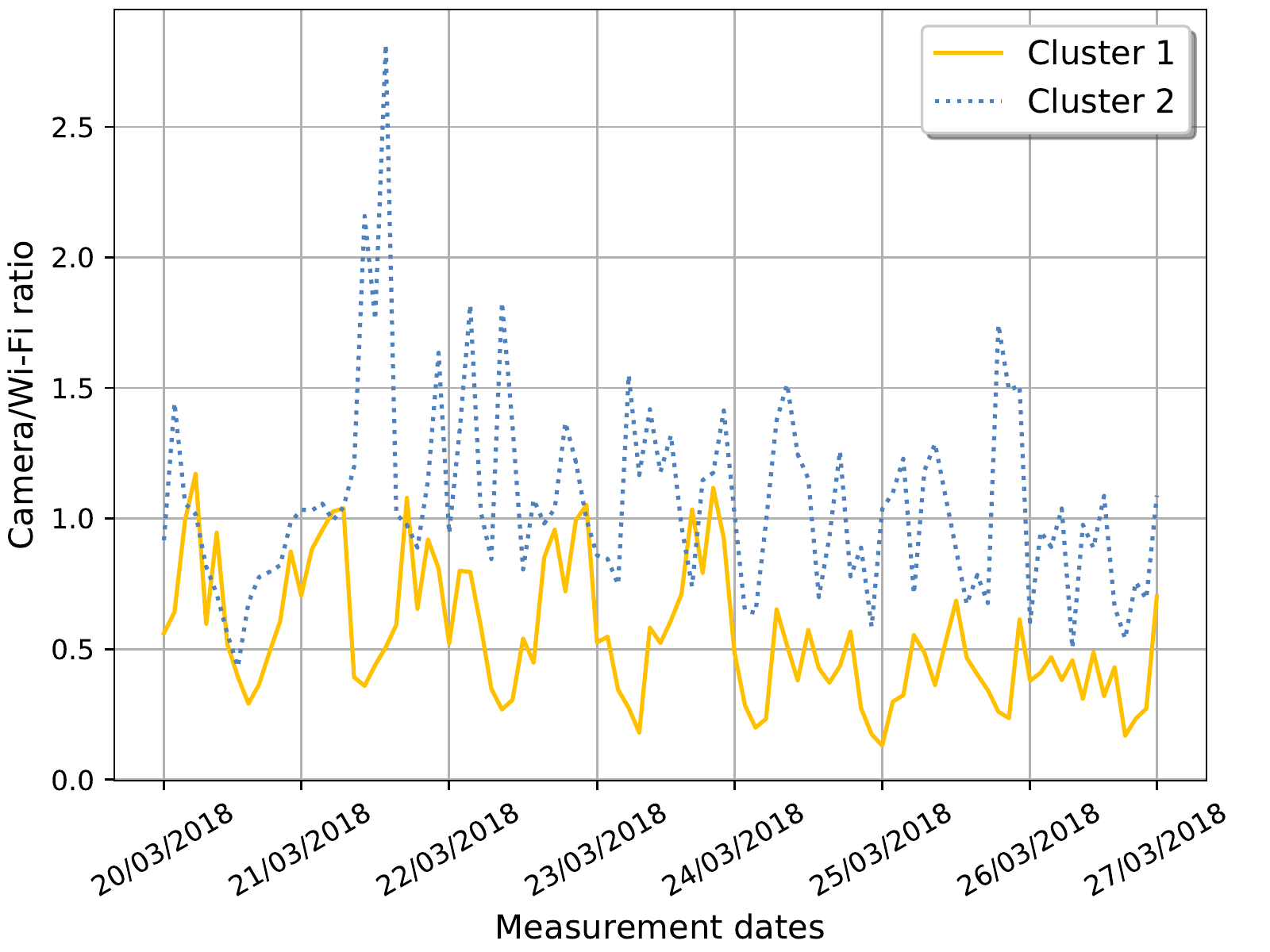} \\
(a) The number of unique Wi-Fi devices detected (hourly). & (b) Hourly changes of the camera/Wi-Fi ratios.  \\
\end{tabular}
\end{framed}
\caption{Weekly measurements from Gold Coast. Cluster 1: Pedestrian ways with heavy pedestrian traffic, Cluster 2: Roads including vehicles and light pedestrian traffic.} \label{Fig:Correlation_Coefficient}
\end{figure*}

\subsection{Pilot Deployment in Santander}
\label{SmSPilotDescription}

\subsubsection{Pilot Setup}
CCLS is deployed in the ``Mercado del Este'' market, a restored symmetric building that contains shops, restaurants, a regional tourist office, and a museum. This building is particularly interesting as it usually receives significant numbers of visitors due to its central location, with exceptionally crowded periods.

The system is composed of 8 devices installed within the market building. These devices include a Wi-Fi interface aimed at detecting surrounding Wi-Fi enabled visitors' devices. Internet connectivity is provided through the Municipality Network, and the devices are powered using Power over Ethernet connected to the market's electrical grid. In addition to the wireless interfaces, half of the devices also include environmental sensors measuring temperature and humidity. Device deployment is carried out with the collaboration and supervision of the municipality and the market managers. Considering the main goal of monitoring people within the market, two parameters are considered in order to get market status snapshots over the time. First, the number of visitors within the market in different time frames and second, the location of the visitors in the different areas of the market.

\subsubsection{Pilot Operation}

Firstly, in order to monitor the visitors within the market, we follow a deterministic approach, in which we consider that a device is inside the building if a minimum of 6 sensor nodes detect it with a certain level of RSSI. In our deployment, this solution is feasible considering the particular symmetric distribution of the building and the location of the sensor nodes, covering the external wall of the building. Secondly, device locations are estimated using the Weighted Centroid Algorithm ~\cite{kosovic2014enhanced}, which provides a reasonable approximation of 5 meters to the ground-truth measurements without any ad-hoc calibration. For the cases that require more precision, these positioning methods are able to introduce less than 2 meters error if the system is calibrated in advance.
Synthesized information including real-time visitor location and detected number of visitors per unit of time is provided through a web portal to the market managers and municipality responsibles. Fig.~\ref{Fig:MercadoDelEsteHeatmap} shows the heat map of the market in a specific moment. Other parameters, such as the visitors' dwell time in different long-term periods, are not analysed due to the privacy safeguards that have to be addressed.

\section{City-Scale Experiments}
\label{Insights}

This section discusses some of the experimental observations from the Gold Coast pilot with CMAS. Specifically, it includes the variance in the crowd estimation for the Wi-Fi sensors and cameras. Our focus in the experimental study is to observe the dynamic changes in the {\em number of unique Wi-Fi devices detected} and the {\em correlation coefficient} (or simply camera/Wi-Fi ratio), which is a dynamic parameter that is computed by the {\em Adaptive Linear Calibration Algorithm}~\cite{Wu-2018-MOBISYS}. The coefficient basically indicates the proportion of the number of people (count-in and count-out events) detected by the camera to the number of devices detected by the Wi-Fi sensors every time interval. We analyze the hourly results for the two clusters, where 5 minute time intervals are aggregated and averaged for 1 hour.

Figure~\ref{Fig:Correlation_Coefficient}-a shows the average number of Wi-Fi devices detected for one-week period. There exists an increased activity in Cluster 1 region especially during Friday (23/03/2018) and the following weekend. This can be due to crowdedness in the shopping street and the beach area contained in this region. Moreover, there is a peak in Saturday that can be due to an event or gathering. Figure~\ref{Fig:Correlation_Coefficient}-b shows the change of the coefficients (ratios). The ratios are computed at the calibration choke points 
(providing near-ground truth for the measurements). 
The hourly ratio is computed such that number of people count-in and count-out events are divided to the number of Wi-Fi probes. First, for Cluster 2 with light traffic, correlation coefficient is mostly (almost all days) higher compared to Cluster 1. Second, correlation coefficient values lie mostly in the range of (0.2, 2), whereas the peak value is about 2.8. This indicates that the results based on Wi-Fi-only measurements are likely to have less accuracy most of the times of the days and the correlation changes throughout the days. Lastly, there exists certain regularity in the correlation from one day to another, which can be learned through a time period and then applied to other time periods where camera is temporarily inactive or removed. On the other hand, as seen in the peak hours of Cluster 2, the ratios do not lie within a narrow range. One reason can be events affecting the volume of pedestrians. Lastly, Fig.~\ref{Fig:Correlation_Coefficient}-b shows relatively higher variance of the coefficient for areas with light pedestrian traffic. Calibration could be necessary for shorter time intervals.

Overall, it is observed that effective use of Wi-Fi sensing and combining them with sensing by stereoscopic cameras produce accurate sensing in large scale for both the heavy and light pedestrian traffic areas. Moreover, the variance between heavy and light traffic shows the usefulness of the clustering approach which treats these regions separately.

\section{Related Work}
\label{RelatedWork}

There are recent studies that focus on understanding of human mobility through IoT devices such as wireless sensors. Jara et al.~\cite{jara2015big} observed the relation between traffic behavior and temperature conditions as a smart city application through deployment of IoT devices in Santander. 
Tong et al.~\cite{tong2017modeling} propose usage of Wi-Fi sensors to understand passenger flows. Evaluation through simulation results shows high accuracy. 
Zhao et al.~\cite{zhao2016urban} survey the recent advances in understanding human mobility in urban environments. The study lists some of the existing urban human mobility datasets collected such as GPS, GSM, Wi-Fi, and Bluetooth traces. Similarly, Zhou et al.~\cite{Zhou-CommMag2018} discuss the topic of human mobility in urban environments and present a taxonomy of crowdsensed input data types and application outcomes such as crowd density and flows within building, and people transportation mode identification (cycling, running, bus riding).
Lastly, Montjoye et al.~\cite{de2013unique} focus on the privacy aspect by analyzing long period Wi-Fi traces and show that 95\% of the individuals can be uniquely identified using spatiotemporal datasets. 



\section{Future Work and Challenges}

The current work focuses on finding insights behind crowd mobility such as detecting crowdedness. However, understanding more complex crowd mobility behaviour in a large-scale city area such as movements of groups (e.g., family) could be helpful for crowd management and enhancing smart mobility in the cities. The collected mobility information can serve as input of human mobility simulations to further study how city dynamics are affected by crowd mobility patterns. With the combination of real mobility dataset in a simulated environment, learning new mobility insights opens up new opportunities for new crowd management strategies (e.g., congestion avoidance, evacuation planning, demand management) that can further improve the public service and safety in smart cities.

In our future developments, the semantic interoperability through ontologies can be leveraged more extensively for cross-infrastructure communication and knowledge sharing. The new advancements of the NGSI protocol by the ETSI Industry Specification Group (ISG) on Context Information Management (CIM) 
are centered around the concepts of linked data. This opens a new horizon where knowledge graphs are shared among various infrastructures and, while their administrators own the produced data, it is still accessible seamlessly and transparently by all actors in the  multi-infrastructure federation. 

\section{Conclusions}
\label{Conclusion}
This article discusses the new advancements towards understanding crowd mobility in smart cities using IoT. While there exist certain limitations, the CMAS and CCLS systems using the smart city platform offer improvements for more efficient crowd management. The pilot studies in Gold Coast and Santander show the capability to fulfill various requirements and share information across stakeholders by leveraging the IoT technologies and infrastructure. 

\parpic{\includegraphics[width=0.2\linewidth,clip,keepaspectratio]{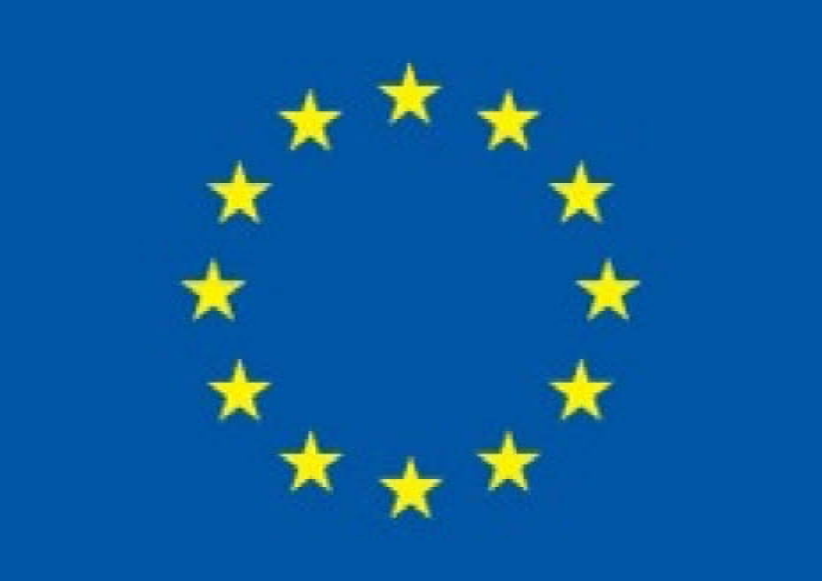}}
\noindent \textbf{Acknowledgment:} The pilot study in Gold Coast is conducted in collaboration with NEC Australia. This work has been partially funded by the Spanish Government (MINECO) under Grant Agreement No. TEC2015-71329-C2-1-R ADVICE (Dynamic Provisioning of Connectivity in High Density 5G Wireless Scenarios) project and by the EU Horizon 2020 Programme under Grant Agreements No. 731993 AUTOPILOT (Automated Driving Progressed by Internet Of Things),  643943 FIESTA-IoT (Federated Interoperable Semantic IoT Testbeds and Applications), and 643275 FESTIVAL (Federated Interoperable Smart ICT Services Development and Testing Platforms) projects and the joint project by NEC Laboratories Europe and Technische Universit\"at Dortmund. The content of this paper does not reflect the official opinion of the Spanish Government or European Union. Responsibility for the information and views expressed therein lies entirely with the authors.

\bibliographystyle{IEEEtran}


\noindent G\"URKAN SOLMAZ (gurkan.solmaz@neclab.eu) is a senior researcher at NEC Laboratories Europe, Germany. His research interests are in mobile computing, human mobility analytics, and Internet of Things.

\noindent FANG-JING WU (fang-jing.wu@tu-dortmund.de) is a junior professor at TU Dortmund. Her research interests are primarily in pervasive computing, wireless communications and networks, and Internet of Things. 

\noindent FLAVIO CIRILLO (flavio.cirillo@neclab.eu) is a research scientist at NEC Laboratories Europe. IoT platforms are his main research focus, with a particular interest to scalability, interoperability, and federation aspects.

\noindent ERN\"O KOVACS (ernoe.kovacs@neclab.eu) is the senior manager for the IoT Research Group, NEC Laboratories Europe. His
context brokering, Cloud-Edge Computing, real-time situation awareness, knowledge extraction, and smart cities.

\noindent JUAN RAM{\'O}N SANTANA (jrsantana@tlmat.unican.es) is a senior research fellow at the University of Cantabria. Among his research interests are wireless sensor networks, M2M communications, and mobile phone application research.

\noindent LUIS S{\'A}CHEZ (lsanchez@tlmat.unican.es) is a professor at the Department of Communications Engineering, University of Cantabria. He is active on IoT-enabled smart cities, meshed networking on heterogeneous wireless scenarios, and optimization of network performance through cognitive networking techniques.

\noindent PABLO SOTRES (psotres@tlmat.unican.es) is a senior research fellow at the University of Cantabria, Spain. He has been involved in several international projects framed under the smart city paradigm, such as SmartSantander; and related to inter-testbed federation, such as Fed4FIRE, Fed4FIRE+ and Wise-IoT.

\noindent LUIS MU{\~N}OZ (luis@tlmat.unican.es) is the head of the Network Planning and Mobile Communications Laboratory at the University of Cantabria, Spain. His research focuses on advanced data transmission techniques, heterogeneous wireless multihop networks, and applied mathematical methods for telecommunications.

\end{document}